\documentclass[11pt,showpacs,showkeys,nofootinbib,preprintnumbers]{revtex4-2}
\usepackage{graphicx}   
\usepackage{color}         
\usepackage{braket}
\usepackage{amsmath}
\usepackage{amssymb}
\usepackage{mathrsfs} 
\usepackage{titlesec}
\usepackage{makecell}
\usepackage{natbib,hyperref}
\usepackage[mathscr]{euscript}
\usepackage{dutchcal}
\usepackage{multirow}

\def \bea {\begin{eqnarray}}
\def \eea {\end{eqnarray}}

\usepackage{mathrsfs}  
\usepackage{bm}

\definecolor{ao-english}{rgb}{0.0, 0.5, 0.0}
\definecolor{cadmiumblue}{rgb}{0.0, 0.42, 0.24}

\newcommand{\nn}{\nonumber}

\begin{document}

\preprint{ECTP-2025-02}
\preprint{WLCAPP-2025-02}
\hspace{0.05cm}

\title{Reevaluation of Inflationary Dynamics in Extended General Relativity with Perturbatively and Tensorially Structured  Conformal Metric}

\email{Corresponding author: a.tawfik@fue.edu.eg; atawfik@bnl.gov}

\author{Swapnil~K.~Singh$^{1}$, Saleh~O.~Allehabi$^{2}$, A.~A.~Alshehri$^{3}$, M.~Nasar$^{4}$ and A.~Tawfik$^{2,5}$}

\affiliation{$^1$BMS College of Engineering, Bangalore, Karnataka, 560019, India} 
\affiliation{$^2$Department of Physics, Faculty of Science, Islamic University of Madinah, Madinah 42351, Saudi Arabia} 
\affiliation{$^3$Department of Science and Technology, University College at Nairiyah, University of Hafr Al Batin (UHB), Nairiyah 31981, Saudi Arabia} 
\affiliation{$^4$Physics Department, Faculty of Science, Benha University, 13511 Benha, Egypt} 
\affiliation{$^5$Ahram Canadian University (ACU), Faculty of Engineering, 12556 Giza, Egypt}

\begin{abstract}

Based on the conventional metric tensor and driven by a nearly constant energy density, 
cosmic inflation, characterized by a remarkably accelerated expansion, was proposed as an early epoch in the Universe. The energy density is typically modeled through a slow-rolling scalar field, whose potential energy dominates the dynamics. This mechanism addresses horizon, flatness, and relic problems, while also generating quantum fluctuations that are stretched to cosmological scales, leading to emergence of primordial curvatures and tensor perturbations. Despite its empirical success, significant questions remain regarding identity of the inflaton, origin of the potential, and role of quantum gravity. A quantum-deformed conformal metric that is both perturbatively and tensorially structured and expanded is employed to reexamine the dynamics of inflation, thus enabling the computation of a range of inflationary observables in presence of quantum-induced corrections. We have established a closed and internally consistent set of analytical formulas for scalar and tensor power spectra, including their spectral tilts, runnings, and the tensor-to-scalar ratio, among other parameters. The quantum corrections appear to provide a clear physical interpretation related to measure scaling and momentum-induced kinetic deformation, which facilitates modifications to the inflationary observables in a controlled and predictive manner. While maintaining the classical limits, these corrections provide a well-defined phenomenological perspective on potential quantum-gravitational structures in the early Universe.
\end{abstract}

\maketitle

\tableofcontents

\section{Introduction}
\label{sec:intrdctn}

There exists a widely shared consensus among theoretical physicists that during the very early stages after the Big Bang, the Universe underwent a rapidly accelerated expansion known as cosmic inflation \cite{gamow2004creation,Barrow1981}. The concept of this inflationary evolution helps to explain several observed features of the Universe, such as large-scale homogeneity, distribution of Cosmic Microwave Background (CMB) radiation \cite{perovic2024cosmic}, and absence of magnetic monopoles \cite{Zhang2024}. While it is presumed that this epoch starts with a swift increase in the scales of the Hubble radius \cite{PhysRevB.21.702,1093mnras195}, significant cosmological disturbances with a scale-invariant spectrum are likely produced, suggesting that the exponential growth is likely confined to the Hubble radius \cite{Liddle:1994yx}. In this context, let us remark that exotic relics like the primordial magnetic monopoles \cite{LINDE1983177}, which could cause remarkable modifications in the large-scale structure, seem not to be fully explained by the inflationary scenario \cite{PhysRevLett.48.1220,PhysRevD.23.347, LINDE1982389}. On the other hand, it is obvious that the rapid expansion might be able to remove the particle horizon, aiding in the resolution of the horizon issue \cite{1980ApJ...241L..59K}. Moreover, the inflationary scenario appears to provide an explanation for the consistency and flatness observed in our Universe \cite{Gonzalez2020}. Also, the inflation is presumed to trigger quantum fluctuations in classical curvature variations \cite{Chibisov:1982nx,Lukash:1980iv}. It is assumed to begin on a small scale and grows over time, thereby it provides an explanation for the large-scale structure of the Universe \cite{Kahn:1984na,Bardeen:1986iq}. While the exact cause of inflation remains unclear, various predictions have been confirmed, with the theoretical field inflaton \cite{guth1998inflationary} regarded as a possible candidate based on observational data \cite{STAROBINSKY198099,TawfikDahab2017}.
All these features have been disclosed by the semiclassical four-dimensional pseudo-Riemann metric. The present manuscript reevaluates the inflationary dynamics through a quantum-deformed metric that is conformally constructed with a perturbative and tensorial expansion.

Cosmic inflation cannot be considered entirely complete unless the relevant context is thoroughly articulated. Following the conclusion of cosmic inflation, the Universe enters a reheating phase, during which the energy stored in the inflaton field is transferred into a hot, dense plasma of particles \cite{German2025}. This process involves complex interactions and decay mechanisms that convert the coherent energy of the inflaton, the hypothetical scalar field, into the degrees of freedom associated with the Standard Model of particle physics \cite{9nn9-bsm9}. In this regard, let us emphasize that during this epoch, the Universe heats up and begins to expand in a manner consistent with the Standard Cosmological Model, setting the initial conditions for nucleosynthesis, structure formation, and CMB \cite{ABDALLA202249}. 

Various theoretical models of inflation attempt to realize these dynamics through different choices of the inflaton potential $V(\phi)$. For instance, Starobinsky inflation \cite{Starobinsky:1979ty,doi:10.1142/9789814317344_0078}, which is based on modifications of gravity and involves a specific form of the potential that naturally leads to a plateau-like region, has gained significant support due to its compatibility with observational data \cite{POZO2024116726}. Other models, such as Natural inflation \cite{Jensen:1986nf}, rely on pseudo-Nambu--Goldstone bosons with sinusoidal potentials \cite{PhysRevLett.65.3233}, while d-brane scenarios incorporate extra-dimensional effects to generate the inflationary phase \cite{DVALI199972}. Despite their differences, all these models depend on maintaining an extended slow-roll regime - where the inflaton field rolls slowly down its potential - to produce enough e-folds of expansion and to generate primordial fluctuations consistent with observations \cite{Gong:2001he,Stewart:2001cd}. 

Empirical constraints from the Planck satellite measurements have provided strong support for the inflationary paradigm, favoring models that predict a scalar spectral index slightly less than unity, indicative of a red tilt in the primordial power spectrum \cite{Enqvist:2013eua,Planck:2013jfk}. Additionally, these observations place stringent upper limits on the tensor-to-scalar ratio, which measures the relative amplitude of primordial gravitational waves to density fluctuations \cite{Mukhanov:2005bu}. These results are consistent with slow-roll inflation occurring on plateau-like potentials, where the potential remains sufficiently flat over a wide range of the inflaton field values \cite{GrootNibbelink:2001qt}. Furthermore, these findings help narrow down the viable models and deepen our understanding of the physics driving the early stages of the Universe.

This study presents a new theoretical framework aimed at enhancing the comprehension of the inflationary period. Utilizing a geometric quantization method, one of the authors (AT) has suggested the geometric quantization of the fundamental metric of General Relativity (GR) \cite{Tawfik:2025wsl,Tawfik:2025icy,Tawfik:2025rel,Tawfik:2025kae,Tawfik:2025ldp,
Tawfik:2024itt,NasserTawfik:2024afw,Farouk:2023hmi,Tawfik:2023ugm,Tawfik:2023rrm}. The proposed quantization ansatz necessitates several generalizations to GR and Quantum Mechanics (QM). Firstly, it integrates gravitational and relativistic fields into QM, thereby modifying the Heisenberg uncertainty principle and subsequently generalizing QM, including among others the Schr{\"o}dinger and Klein--Gordon equations. By utilizing the Relativistic Generalized Uncertainty Principle (RGUP) \cite{Benczik:2002tt,Todorinov:2020jtq,phdthesisXun,Tawfik:2023onh,Tawfik:2023orl}, modified quantum operators were derived, the most significant of which is the momentum operator. Secondly, the quantization ansatz suggests a generalization of Riemann geometry within GR, enabling a transition from four-dimensional spacetime to phase-space spacetime. The geometry of phase-space permits the incorporation of quantum modifications into the associated metric. This integration is achievable through the implementation of a theory about the kinematics of quantum particle movement in curved spacetime. The metric derived from the extended Finsler and Hamilton structures, that includes quantum corrections, is then transformed back into Riemann geometry. To this end, it is assumed that the line element in both geometries is the same. This may be an approximate approach, but in the absence of other alternatives, it could be tolerated. The transition to Riemann geometry is dictated by the postulates of GR.

This paper presents a thorough examination of the impact that a quantum-modified metric has on the inflationary epoch of the early Universe. We investigate the theoretical consequences of the quantum corrections. The study seeks to reveal how they may influence the dynamics of cosmic inflation, encompassing the expansion rate, the creation of primordial fluctuations, and the overall development of the Universe during this pivotal phase. Various models and mathematical constructs that include the notion of a quantized metric are reviewed. Their alignment with established inflationary theories and empirical data is evaluated. The results indicate that the inclusion of quantum geometric effects within the inflationary framework could yield new insights, potentially clarifying certain anomalies or characteristics observed in the inflationary dynamics. In conclusion, the paper asserts that this methodology enhances our comprehension of the complex interplay between quantum gravity - an essential field of physics aiming to merge GR with QM - and the large-scale structure of the Universe. This research paves the way for new explorations, highlighting the significance of quantum geometric revisions in cosmology and the ongoing quest to unify the fundamental forces of nature through a quantum theory of gravity.

The present script is structured as follows: The formalism is presented in Section \ref{sec:frmlsm}. This begins with a concise review of the geometric quantization of the the fundamental metric in Section \ref{sec:qMtrTnsr}. Additionally, the approximating method for the quantum-deformed metric is examined in Section \ref{sec:PrtrbMetric}. The perturbative and tensorial approximate structure, developed to articulate the quantum-deformed metric, is derived in Section \ref{sec:PrtrbMetric}. This enables the presentation of analytical results in Section \ref{sec:AnltcReslt}. The details of the quantum first-order expansion, background evolution, and spectrum are elaborated in Section \ref{sec:qFOexp}. The power-law inflation, Starobinsky Inflation, d-brane (Coulomb) inflation, and natural inflation, along with their modifications due to the proposed perturbative tensorial structure incorporated in the fundamental metric, are summarized in Sections \ref{sec:QPL-infl}, \ref{sec:StarobinskyInfl}, \ref{sec:dBrane}, and \ref{sec:NatInfl}, respectively. The numerical results are presented in Section \ref{sec:NmrcReslt}. Section \ref{sec:conclusions} is dedicated to the concluding remarks and future perspectives.

\section{Formalism}
\label{sec:frmlsm}

A brief review of the quantum-induced revisions on the fundamental metric and, as a result, the Einstein tensor is provided. Detailed information about this approach can be found in ref. \cite{Tawfik:2025wsl}. The geometric quantization ansatz for GR, inspired by the Relativistic Generalized Uncertainty Principle (RGUP) \cite{phdthesisXun,Todorinov:2020jtq,Tawfik:2024gxc}, is reformulated. We commence our discussion in Section \ref{sec:qMtrTnsr} with the derivation of the quantum-deformed metric \cite{Tawfik:2025wsl}.

\subsection{A Brief Review of Quantum-Deformed Metric}
\label{sec:qMtrTnsr}

The RGUP suggests to modify momentum operator \cite{phdthesisXun,Todorinov:2020jtq,Tawfik:2024gxc}. Thus, in phase-space geometry, the corresponding Hamilton metric  $\tilde{g}^{(pp)}_{\mu \nu}$ can be derived as
\begin{equation}
g_{\mu \nu}(x,p) = \frac{1}{2} \frac{\partial^{2}}{\partial p_{0}^{\mu} \partial p_{0}^{\nu}} F(x, \phi^{2}(p_{0})p),
\label{eq:Hessian1}
\end{equation}
where the function $\phi(p_{0})$ is deduced from the RGUP approach, $\phi(p_{0})=1+\beta\,p_{0}^{\rho}p_{0\rho}$ with $beta$ being the RGUP parameter that can be determined either empirically or observationally \cite{Diab:2020jcl}. Here, $\rho$ serves as a dummy index and $p_{0}$ represents the auxiliary four-momentum which is noncommutative with $x_{0}$, the auxiliary four-vector distance. In this context, the differentiation with respect to the momentum variables is consistent, as $F(x,p)$ adheres to Hamilton structure, which can be obtained from the Finsler structure $F(x,\dot{x})$ at least by considering the homogeneity of positive $F(x,\dot{x})$ \cite{Hamilton1982,Albuquerque:2023icp,PhysRevD.92.084053,Bubuianu:2018wry}. For instance, for a positive mass $m$ of a quantum particle, $m F(x,\dot{x})=F(x,m\dot{x})=F(x,p)$ holds true. The resulting metric $\tilde{g}_{\mu \nu}(x,p)$ transforms as a rank-$(0,2)$ tensor on the base manifold and corresponds to the standard pseudo-Riemann metric when $ \phi(p_0) \rightarrow 1$. Consequently, the Finsler function, for example, becomes quadratic and thus defines Riemann geometry. The requirement for strong convexity in the Hamilton structure \cite{Mebrate2024} necessitates that the Hessian in Eq.~\ref{eq:Hessian1} must be positive definite, which guarantees the reversibility of Legendre mapping and the symmetry condition of $\tilde{g}_{\mu \nu}$. In this regard, let us emphasize that the quantum-induced modifications are presumed to arise through higher-order terms in the scalar $p_{0}^{\rho} p_{0 \rho}$, thereby modifying the metric whenever $\phi(p_0) \neq 1$.

As $\phi(p_0)$ is zero-homogeneous in $p_0$, the function $F(x, \phi(p_{0})p)$ is homogeneous of degree two and Euler theorem leads to
\bea
p_{0}^{\mu} \frac{\partial}{\partial p_{0}^{\mu}} F(x, \phi^{2}(p_{0})p) &=& 2 \phi^{2}(p_{0})\, F^{2}(x, p).
\eea 
Upon second differentiation, the bilinear form reads 
\begin{equation}
\phi^{2}(p_{0})\, F^{2}(x, p) = g_{\mu \nu}(x, p)\, \dot{p}_{0}^{\alpha} \dot{p}_{0}^{\beta},
\label{eq:F2gab}
\end{equation}
follows along any curve $\zeta \mapsto p_{0}(\zeta)$ in momentum space. This relation should not be interpreted as a Finslerian length on configuration-space displacements; rather, it provides the second-order expansion of $\phi^{2}(p_{0}) F^{2}(x, p)$ along momentum-space directions and is a standard construction in Finsler/Hamilton geometry. 

For the sake of simplicity, let us assume Klein metric on $\mathbb{R}^{n}$ \cite{Mo2006,Klein:1926tv}, which belongs to the set of simplest Finsler metrics and on Hamilton manifold\footnote{As previously mentioned, a Finsler structure can be converted into a Hamilton structure through either the homogeneity property or the Legendre transformation \cite{Hamilton1982,Albuquerque:2023icp,PhysRevD.92.084053,Bubuianu:2018wry}.} can be expressed as  
\begin{equation}
F(x, p) = \left[\frac{|p|^{2} - |x|^{2}\, |p|^{2} + \langle x, p\rangle^{2}}{1 - |x|^{2}}
\right]^{1/2},
\label{eq:FKlein}
\end{equation}
where $|\cdot|$ gives the Euclidean norm and $\langle\cdot,\cdot\rangle$ the corresponding inner product. It apparent that 
Legendre inversion retrains the associated Finsler structure \cite{Hamilton1982,Albuquerque:2023icp,PhysRevD.92.084053,Bubuianu:2018wry}.  

Combining the Hessian matrix in Hamilton geometry, Eq. \eqref{eq:Hessian1}, with the coordinates $(x,p) \equiv (x^{\mu}, p_{\nu})$ results in quantum-deformed metric, which can be translated to four-dimensional Riemann geometry\footnote{The conversion of the Finsler/Hamilton metric to the pseudo-Riemann metric is fundamentally based on the premise that the line elements in both geometrical types are the same. This may introduce an additional source of approximation that could be addressed elsewhere, Section \ref{sec:PrtrbMetric}.}
\bea
\tilde{g}_{\mu \nu}(x,p) &=& \left[\phi^{2}(p_{0}) + 2 \frac{\kappa}{(p_{0}^{0})^{2}}\, F^{2}(x, p) \right] \left[1 + \frac{\dot{p}_{0}^{\alpha} \dot{p}_{0}^{\beta}}{{\mathscr F}^{2}} \left(1 + 2 \beta p_{0}^{\rho} p_{0 \rho}\right)\right] g_{\mu \nu}(x) \nn \\
&+& \left[\frac{dx_{0}^{\alpha}}{d\zeta} \frac{dx_{0}^{\beta}}{d\zeta} +
\left(1 + 2 \beta p_{0}^{\rho} p_{0 \rho}\right) \frac{dp_{0}^{\alpha}}{d\zeta}  \frac{dp_{0}^{\beta}}{d\zeta} \right] d_{\mu \nu}(x),
\label{eq:gmunuQ2}
\eea
where $\kappa = \beta/(p_0^0)^2$ and an overdot denotes differentiation with respect to the common curve parameter $\zeta$. No index is ever repeated improperly. The term $\dot{p}_{0}^{\alpha} \dot{p}_{0}^{\beta}$ represents the squared proper four-force. In this respect, ${\mathscr F}$ is the maximal proper force, which can be related to the maximal proper acceleration ${\mathscr A} = (c^{7}/\hbar G)^{1/2}$ \cite{Caianiello:1981jq,caianiello1984maximal,Brandt:1988sh}. 

A coarse approximation of the quantum-deformed metric is discussed in Section 
\ref{sec:PrtrbMetric}.

\subsection{Approximating Approach to Quantum-Deformed Metric}
\label{sec:PrtrbMetric}

The resulting metric $\tilde{g}_{\mu\nu}(x,p)$ is a symmetric rank-$(0,2)$ tensor with correct physical dimensions and reduces to the conventional pseudo-Riemann metric in the limit that $\beta \to 0$ and $\kappa \to 0$. The first term in \eqref{eq:gmunuQ2} encodes momentum-space curvature and quantum corrections, while the second term incorporates acceleration-induced deformations inherited from the tangent-bundle geometry. Note that $d_{\mu \nu}(x)$ is a deformation tensor projecting onto tangent-bundle directions and is smooth, symmetric, and dimensionless. Due to complexity in being  expressed in terms of $g_{\mu \nu}(x)$, the quantities within the square brackets aren't evaluated. The mathematical difficulties associated with expressing $d_{\mu \nu}(x)$, Eq. \eqref{eq:gmunuQ2}, as a conformal transformation of $g_{\mu \nu}(x)$ might be addressed elsewhere \cite{Mubaidin2026,Singh:2026dyn}. To ensure adherence to the scope of this manuscript, we presume that the vanishing of 
$d_{\mu \nu}(x)$ might be tolerate, particularly in relation to the coarse approximation where the entire Sasaki metric is still represented by just one $(pp)$-component $g_{\mu \nu}(x,p)$.

Starting from the entire quantum-induced revisited metric $\tilde{g}_{\mu \nu}(x,p)$, Eq.~\eqref{eq:gmunuQ2}, let us assume that  the perturbative regime where the quantum-gravity parameters $\beta$ and $\kappa$ are small is a suitable scale. Here the higher-order powers of $\beta$ and $\kappa$ can be neglected and one obtains an approximate, conformally deformed metric on the four-dimensional spacetime \cite{Tawfik:2025wsl,Tawfik:2025icy,Tawfik:2025rel,Tawfik:2025kae,Tawfik:2025ldp,
Tawfik:2024itt,NasserTawfik:2024afw,Farouk:2023hmi,Tawfik:2023ugm,Tawfik:2023rrm}.  Retaining only the leading-order contributions, the quantum-induced metric can be truncated as
\begin{equation}
\tilde{g}_{\mu\nu}(x, p) = \left(\phi^{2}(p_0) + 2 \frac{\kappa}{(p_{0}^{0})^{2}} F^{2}(x, p) \right) \left[1 + \frac{\dot{p}_{0}^{\alpha} \dot{p}_{0}^{\beta}}{{\mathscr F}^{2}} \left(1 + 2 \beta p_{0}^{\sigma} p_{0 \sigma} \right) \right] g_{\mu \nu}(x),
\label{eq:gmunuQ2b}
\end{equation}
where $\phi(p_{0})$ and $\kappa$ are dimensionless and $p_{0}^{\rho} p_{0 \rho}$ is the Lorentzian contraction of the undeformed momentum. It is convenient to collect all multiplicative corrections into a single scalar conformal factor $C(x,p)$ apparently defined as
\begin{equation}
C(x,p) = \left(\phi^{2}(p_0) + 2 \frac{\kappa}{(p_{0}^{0})^{2}} F^{2}(x, p) \right) \left[1 + \frac{\dot{p}_{0}^{\alpha} \dot{p}_{0}^{\beta}}{{\mathscr F}^{2}} \left(1 + 2 \beta p_{0}^{\sigma} p_{0 \sigma} \right) \right].
\end{equation}
Since all terms in $C(x,p)$ are dimensionless, the conformal factor itself is a dimensionless scalar field in $(x,p)$ space. Therefore, the quantum-induced metric can be written as a simple conformal rescaling of the classical background metric,
\begin{equation}
\tilde{g}_{\mu \nu}(x,p) = C(x, p)\, g_{\mu \nu}(x).
\label{eq:tildegalphabeta1}
\end{equation}

Regarding the {\it approximate} formulation of the quantum-deformed metric, the implications of the perturbative approach are detailed in Section \ref{sec:PrtrbMetric} .

\subsection{Perturbative and Tensorial Approximate Structure of Quantum-Deformed Metric}
\label{sec:PrtrbMetric}

Let us now recall that in a fully quantum formulation, Eq. \eqref{eq:gmunuQ2}, momenta $p_{0}^{\alpha}$ and relevant derivatives $\dot{p}_{0}^{\alpha}$ are quantum operators. For practical implications, such as cosmology and astrophysics, it is customary to evaluate expectation values of the quantum operators utilizing mean-field approximation, for instance. One then replaces operator-valued quantities by their ensemble averages, for example $\langle p_{0}^{\alpha}\rangle$ and $\langle \dot{p}_{0}^{\alpha}\rangle$, while preserving the functional dependence of the conformal coefficient $C(x,p)$. This yields an effective $\langle C(x,p)\rangle$ that encodes the leading quantum corrections in a semiclassical or canonical quantum description \cite{Tawfik:2024itt,NasserTawfik:2024afw}.

To analyze the structure of the quantum-corrected geometry on the underlying Riemann manifold, it is useful to rewrite the conformal deformation coefficient in a perturbative and tensorial form. To this end, we introduce a first-order {\it deformation} tensor $A_{\mu}^{\alpha}(x,p)$ and accordingly find that
\begin{equation}
\tilde{g}_{\mu \nu}(x,p) = \left[\delta_{\mu}^{\alpha} + A_{\mu}^{\alpha}(x,p)\right] g_{\alpha\nu}(x),
\label{eq:metric_with_A}
\end{equation}
where $A_{\mu}^{\alpha}$ is a small and dimensionless tensor encoding the deviation from the classical metric. In the approximation considered so far, this tensor might be taken as
\bea
A_{\mu}^{\alpha}(x,p) \simeq \mathcal{O}(\beta, \kappa),
\eea
where finite $\beta$ and $\kappa$ are  parameters associated with quantum ingredients. As discussed earlier, $\kappa = \beta/(p_0^0)^2$ with $\beta$ being the RGUP parameter that should not be interchanged with the index $\beta$. It is obvious that all indexes in $A_{\mu}^{\alpha}(x,p)$ can be raised and lowered by $g_{\mu \nu}$. A convenient parametrization of the deformation tensor, motivated by Eq.~\eqref{eq:gmunuQ2b}, is proposed as
\begin{equation}
A_{\mu}^{\alpha}(x, p) = \Xi(x, p) \frac{\dot{p}_{0}^{\alpha}  \dot{p}_{0 \mu}}{{\mathscr F}^{2}},
\end{equation}
where the scalar function $\Xi(x, p)$ combines the dependence on the quantum revision, such as RGUP, as well as on the curvature parameters. In terms of leading order with respect to $\beta$ and $\kappa$, this function can be parameterized as
\begin{equation}
\Xi(x, p) = \left[1 + 4 \beta p_{0}^{2}\right] + \left[1 + 3 \beta p_{0}^{2}\right] \frac{2 \kappa}{(p_{0}^{0})^{2}} F^{2}(x, p),
\end{equation}
where $p_{0}^{2}=p_{0}^{\rho} p_{0 \rho}$ and $F^{2}(x,p)$ representing the squared Finsler norm. The latter can be for the sake of simplicity identified with Klein metric \cite{Klein1910,Klein:1926tv}. This parametrization shows explicitly that the deformation vanishes when $\beta \to 0$ and $\kappa \to 0$.

The next step is to derive the inverse metric, $\tilde{g}^{\mu \nu}$, to first order in the deformation. By definition, the inverse metric satisfies $\tilde{g}^{\mu \rho} \tilde{g}_{\rho \nu} = \delta^{\mu}_{\nu}$. The goal is then to formulate a perturbative expansion of the form
\begin{equation}
\tilde{g}^{\mu \nu} = g^{\mu \nu} + \delta g^{\mu \nu}, \label{eq:gupmunu}
\end{equation}
where $\delta g^{\mu \nu}$ is of order $\mathcal{O}(A)$ and all index manipulations are performed with the background metric $g_{\mu \nu}$ and its inverse $g^{\mu \nu}$.  As in Eq.~\eqref{eq:metric_with_A},  Eq.~\eqref{eq:gupmunu} can be reexpressed to the first order in $A_{\mu}^{\alpha}$, therefore
\begin{equation}
\left(g^{\mu \rho} + \delta g^{\mu \rho}\right) \left[\delta_{\rho}^{\sigma} + A_{\rho}^{\sigma}(x,p)\right] g_{\sigma\nu} = \delta^{\mu}_{\nu}.
\end{equation}
Neglecting terms of higher order and using $g^{\mu\rho}g_{\rho\nu}=\delta^{\mu}_{\nu}$, one obtains
\begin{equation}
g^{\mu\rho} A_{\rho}^{\sigma}(x,p) g_{\sigma\nu} + \delta g^{\mu\rho}g_{\rho\nu}=0,
\end{equation}
which implies that
\begin{equation}
\delta g^{\mu\nu} = - g^{\mu\rho} A_{\rho}^{\nu}(x,p)  = - \Xi(x,p) \frac{\dot{p}_{0}^{\alpha} \dot{p}_{0 \beta}}{{\mathscr F}^{2}}.
\end{equation}
Thus, the inverse metric to first order in the deformation is given as 
\begin{equation}
\tilde{g}^{\mu \nu} = g^{\mu \nu} -
\Xi(x,p) \frac{\dot{p}_{0}^{\alpha} \dot{p}_{0 \beta}}{{\mathscr F}^{2}}.
\label{eq:inv_expansion2}
\end{equation}

This perturbative inversion preserves the correct tensorial rank and ensures that
\begin{equation}
\tilde{g}^{\mu \rho} \tilde{g}_{\rho \nu}
 = \delta^{\mu}_{\nu} + \mathcal{O}(A^{2}),
\end{equation}
so that the error is of second order in the quantum corrections. Consequently, the dimensions of $\tilde{g}_{\mu \nu}$ and $\tilde{g}^{\mu \nu}$ are the same as those of the background metric and its inverse, respectively. Furthermore, the overall construction is dimensionally consistent.

The perturbative expansion remains valid as long as $|\Xi(x,p)| \ll 1$, which guarantees that quantum corrections are genuinely small compared to the classical contributions. Under this condition the indexes can be raised and lowered using the background metric $g_{\mu \nu}$ without loss of consistency, and the geometric objects constructed from $\tilde{g}_{\mu \nu}$ and $\tilde{g}^{\mu \nu}$, such as the Christoffel symbols and curvature tensors, retain the correct transformation properties.  

The section that follows, Section \ref{sec:AnltcReslt}, builds on this perturbative tensorial framework to analytically derive the quantized Christoffel symbols and Levi--Civita connections associated with the quantum-deformed metric.

\section{Analytical Results} 
\label{sec:AnltcReslt}

Various types of quantum-corrected inflationary observable are analyzed, at fixed e-folds. The quantum first-order expansion, background evolution, and spectrum in flat Friedmann--Lemaitre--Robertson--Walker (FLRW) background are discussed in section \ref{sec:qFOexp}.

\subsection{Quantum First-Order Expansion, Background Evolution, and Spectrum}
\label{sec:qFOexp}

For reasons of simplicity, let us assume  spatially flat FLRW metric,
\bea
ds^{2} = - dt^{2} + a(t)^{2} \delta_{ij} dx^{i} dx^{j}, 
\eea
where $a(t)$ is the scale factor, 
\bea
N(t) \equiv \ln a(t), \qquad
\frac{d}{dN(t)} = \frac{1}{H(t)} \frac{d}{dt},
\eea
where $H(t) \equiv \dot{a}(t)/a(t)$, the Hubble parameter. The proposed geometric quantization of Finsler structure induces first-order deformation of the inverse metric and volume element, respectively, 
\bea
\tilde g^{\mu \nu} &=& g^{\mu \nu} - \Xi(x,p) \frac{\dot{p}_{0}^{\alpha} \dot{p}_{0 \beta}}{\mathcal F^{2}} + \mathcal O(\Xi^{2}(x,p)),
\label{eq:QC_metric_inv_def_final}
\\
\sqrt{-\tilde g} &=& \sqrt{-g} \left(1 + \frac{1}{2} \Xi(x,p) \frac{\dot{p}_{0}^{\lambda} \dot{p}_{0 \lambda}}{\mathcal F^{2}}\right) + \mathcal O(\Xi^{2}(x,p)).
\label{eq:QC_sqrtg_def_final}
\eea
In homogeneous backgrounds the deformation enters through two scalar contractions,
\begin{equation}
E(N(t)) \equiv \Xi(x,p) \frac{\dot{p}_{0}^{\lambda} \dot{p}_{0 \lambda}}{\mathcal F^{2}}, \qquad
U(N(t)) \equiv \Xi(x,p) \frac{(\dot{p}_{0}^{0})^{2}}{\mathcal F^{2}},.
\label{eq:EU_defs_final}
\end{equation}
where $|E|, |U| \ll 1$. 
There is no algebraic relationship between $E$ and $U$ to be imposed. As a special case, it is presumed that $E \simeq -U$ holds only under an additional assumption of comoving frame.

A canonical inflaton minimally coupled to the deformed geometry has action that expands consistently to $\mathcal O(\Xi(x,p))$
\begin{equation}
\tilde{S}_{\varphi} = - \int d^{4}x\, \sqrt{-g}
\left[\frac{1}{2} F^{\mu \nu} \partial_{\mu}\varphi\, \partial_{\nu} \varphi + \left(1 + \frac{E}{2}\right) V(\varphi) \right]  \mathcal O(\Xi^{2}(x,p)),
\label{eq:phi_action_final}
\end{equation}
where the effective kinetic tensor is given as
\begin{equation}
F^{\mu \nu} = \left(1 + \frac{E}{2}\right) g^{\mu \nu} - \Xi(x,p) \frac{\dot{p}_{0}^{\alpha} \dot{p}_{0 \beta}}{\mathcal F^{2}}.
\label{eq:Fmunu_final}
\end{equation}
Varying Eq. \eqref{eq:phi_action_final} yields Euler--Lagrange equation
\begin{equation}
\nabla_{\mu}\! \left(F^{\mu \nu}\partial_{\nu} \varphi\right) = \left(1 + \frac{E}{2}\right) V'(\varphi).
\label{eq:KG_general_final}
\end{equation}
For a homogeneous inflaton $\varphi=\varphi(t)$, only the zero component of $F^{\mu\nu}$, i.e. $\mu=\nu=0$ contributes to the kinetic operator. Using $g^{00}=-1$ and the definition in Eq. \eqref{eq:EU_defs_final}, one gets 
\begin{equation}
F^{00} = \left(1 + \frac{E}{2}\right) g^{00}
- \Xi(x,p) \frac{(\dot{p}_{0}^{0})^{2}}{\mathcal F^{2}} = - \left(1 + \frac{E}{2}+U\right).
\label{eq:F00_final}
\end{equation}
Since $\nabla_{\mu} X^{\mu}=(\sqrt{-g})^{-1} \partial_{\mu}(\sqrt{-g} X^{\mu})$ and $\sqrt{-g}=a^{3}(t)$, then Eq. \eqref{eq:KG_general_final} can be reduced to
\begin{equation}
\frac{1}{a^{3}(t)} \frac{d}{dt}\! \left[a^{3}(t)\left(1 + \frac{E}{2} + U\right) \dot{\varphi}\right] = - \left(1+\frac{E}{2}\right) V'(\varphi).
\label{eq:KG_FLRW_final}
\end{equation}
Expanding the time derivative leads to
\begin{equation}
\left(1 + \frac{E}{2} + U\right) \ddot{\varphi} + 3 H \left(1 + \frac{E}{2} + U\right) \dot{\varphi} + \left(\frac{\dot{E}}{2} + \dot{U}\right) \dot{\varphi}  = - \left(1 + \frac{E}{2}\right) V'(\varphi).
\label{eq:KG_expanded_final}
\end{equation}
Let us now impose a generalized slow-roll regime,
\begin{equation}
|\ddot{\varphi}| \ll 3 H| \dot{\varphi}|, \qquad \left|\frac{\dot{E}}{H}\right| \ll 1,
\qquad \left|\frac{\dot{U}}{H}\right| \ll 1,
\label{eq:SR_conditions_final}
\end{equation}
so that the first and third terms on left hand side of Eq. \eqref{eq:KG_expanded_final} lead to the Hubble friction term. Retaining leading contributions gives
\begin{equation}
3 H \left(1 + \frac{E}{2} + U\right) \dot{\varphi}  \simeq - \left(1 + \frac{E}{2}\right) V'(\varphi).
\label{eq:phidot_eq_final}
\end{equation}
Solving perturbatively to first order in $(E, U)$ and using $(1 + x)^{-1} = 1 - x + \mathcal{O}(x^{2}(x,p))$ yields
\bea
\dot{\varphi} & \simeq & - \frac{V'}{3 H}\,
\frac{1 + \frac{E}{2}}{1 + \frac{E}{2} + U}
= - \frac{V'}{3 H} \left(1 - U\right) + \mathcal{O}(\Xi^{2}(x,p)),
\label{eq:phidot_solution_final}
\eea
showing that the isotropic deformation proportional to $E$ cancels from the background inflaton velocity at $\mathcal{O}(\Xi(x,p))$.

The leading slow-roll Friedmann equation is taken as
\begin{equation}
3 M_{\rm Pl}^{2} H^{2} \simeq V(\varphi),
\label{eq:Friedmann_SR_final}
\end{equation}
where $M_{\rm Pl}$ is the Planck mass. 
Since inserting $\mathcal{O}(\Xi(x,p))$ corrections to $H^{2}(t)$ into Eq.  \eqref{eq:phidot_solution_final} would generate only $\mathcal{O}(\Xi^{2}(x,p))$, it also changes $\dot{\varphi}$ in all first-order observable.

The e-fold integral between horizon exit ($\ast$) and the end of inflation ($\rm end$) is given as 
\bea
N &\equiv & \int_{t_{\ast}}^{t_{\rm end}} H\, dt
= 
\simeq 
\frac{1}{M_{\rm Pl}^{2}} \int_{\varphi_{\rm end}}^{\varphi_{\ast}} \frac{V}{V'}\, d\varphi + \frac{U_{\star}}{M_{\rm Pl}^{2}} \int_{\varphi_{\rm end}}^{\varphi_{\ast}} \frac{V}{V'}\, d\varphi,
\label{eq:N_master_final}
\eea
where the adiabatic approximation $U(N) \simeq U_{\ast}$ to first order is utilized.

For scalar perturbations, the quadratic action for the Mukhanov--Sasaki variable \cite{Gomar:2014faa,CastelloGomar:2016rjj} remains of the standard form but with a modified normalization \cite{Gomar:2014faa,CastelloGomar:2016rjj}. By defining
\begin{equation}
\delta_{K}(N) \equiv \frac{E(N)}{2} + U(N),
\label{eq:deltaK_def_final}
\end{equation}
the effective field is then given as 
\begin{equation}
z_{\rm eff} = z \left(1 + \frac{\delta_{K}}{2}\right), \qquad z \equiv \frac{a \dot{\varphi}}{H}.
\label{eq:zeff_def_final}
\end{equation}
Accordingly, the mode equation reads 
\begin{equation}
v_{k}'' + \left(k^{2} - \frac{z_{\rm eff}''}{z_{\rm eff}}\right) v_{k} = 0,
\label{eq:MS_equation_final}
\end{equation}
where one prime denotes $d/d\tau$ in conformal time. Evaluating the spectrum at horizon exit $k_{\ast} = a_{\ast} H_{\ast}$ results in
\begin{equation}
P_{\mathcal R}(k_{\ast}) \simeq \frac{1}{24 \pi^{2} M_{\rm Pl}^{4}} \frac{V_{\ast}}{\epsilon_{V\ast}} \left(1 - \frac{E_{\ast}}{2} - U_{\ast}\right) + \mathcal{O}(\Xi^{2}(x,p)),
\label{eq:PR_final}
\end{equation}
where $\epsilon_{V} = \frac{M_{\rm Pl}^{2}}{2} (V'/V)^{2}$ and $\epsilon$ is slow-roll parameter that measures how fast the Universe is departing from exact exponential expansion. Furthermore, $\epsilon$ quantifies the relative change of the Hubble parameter $H$ during inflation and also determines  whether the expansion is accelerating.

The response of tensor perturbations is limited to the rescaling of the measure at the retained order. Equivalently, one may define an effective scale factor $a_{\rm eff}=a(1 + E/2)$ at $\mathcal{O}(\Xi(x,p))$, yielding
\begin{equation}
P_{h}(k_{\ast}) \simeq \frac{2}{\pi^{2}} \frac{H_{\ast}^{2}}{M_{\rm Pl}^{2}} \left(1 + E_{\ast}\right) + \mathcal{O}(\Xi^{2}(x,p)).
\label{eq:Ph_final}
\end{equation}
Therefore, the tensor-to-scalar ratio is given as
\bea
r & \equiv & \frac{P_{h}}{P_{\mathcal{R}}}
\simeq 16 \epsilon_{V\ast} \left(1 + \frac{3}{2} E_{\ast} + U_{\ast}\right)
+ \mathcal{O}(\Xi^{2}(x,p)).
\label{eq:r_master_final}
\eea

The spectral tilts and running follow by differentiating with respect to $\ln k$. Using the exact identity
\begin{equation}
\frac{d \ln k}{dN} = 1 - \epsilon_{H},
\qquad \epsilon_{H} \equiv - \frac{d\ln H}{dN}, \qquad \frac{d}{d \ln k} = \frac{1}{1 - \epsilon_{H}} \frac{d}{dN},
\label{eq:dlnk_dN_final}
\end{equation}
and retaining terms consistently to first order in $\Xi(x,p)$ and first order in slow-roll, we derive, Appendix \ref{app:physical_interpretation},
\begin{equation}
P_{\mathcal R} = P_{\mathcal R}^{(0)}\left(1 + \Delta_{s}\right), \qquad \Delta_{s}(N) \equiv - \frac{E(N)}{2} - U(N),
\label{eq:Deltas_def_final}
\end{equation}
so that the derivative of $\ln P_{\mathcal R}$ with respect to $d\ln k$ is equivalent to $n_{s} -1$, 
\bea
n_{s} - 1 &=& 
\left(n_{s}^{(0)} - 1\right) - \left(1 + \epsilon_{H}^{(0)}\right)\left(\frac{E'}{2} + U'\right) + \mathcal{O}(\Xi^{2}(x,p)),
\label{eq:ns_general_final}
\eea
where one prime here represents $d/dN$ and $\epsilon_{H}^{(0)}$ is the conventional value of $\epsilon_{H}$.

Similarly, for the tensor spectrum we assume that $P_{h} = P_{h}^{(0)}(1 + \Delta_{t})$. Then
\begin{equation}
\Delta_{t}(N(t)) \equiv E(N(t)).
\label{eq:Deltat_def_final}
\end{equation}
Thereby, the derivative of  $\ln P_{h}$ with respect to $d\ln k$ is equivalent to $n_{t}$,
\bea
n_{t} 
&=& - 2 \epsilon_{H}^{(0)} + \left(1 + \epsilon_{H}^{(0)}\right) E' + \mathcal{O}(\Xi^{2}(x,p)).
\label{eq:nt_general_final}
\eea
The running $\alpha_{s} \equiv \frac{dn_{s}}{d\ln k}$ and running-of-running $\beta_{s} \equiv \frac{d\alpha_{s}}{d\ln k}$ (retaining only terms linear in $\Xi(x,p)$ and dropping products beyond the stated accuracy) are expressed as follows:
\bea
\begin{array}{ccl}
\alpha_{s} &=& \alpha_{s}^{(0)}-\left(\frac{E''}{2}+U''\right)+\mathcal O(\Xi^{2}(x,p)), \\
\beta_{s} &=& \beta_{s}^{(0)} - \left(\frac{E'''}{2} + U'''\right) + \mathcal O(\Xi^{2}(x,p)).
\end{array}
\label{eq:alpha_beta_general_final}
\eea

Eliminating $\epsilon_{H}^{(0)}$ between Eqs. \eqref{eq:nt_general_final} and Eq. \eqref{eq:alpha_beta_general_final} and by applying $r = 16 \epsilon_{H}^{(0)}(1 + \frac{3}{2}E + U) + \mathcal{O}(\Xi^{2}(x,p))$, the first-order corrected consistency relationship is then given as
\begin{equation}
r = - 8 n_{t} \left[1 + \frac{3}{2} E + U - \frac{E'}{2 \epsilon_{H}^{(0)}}\right] + \mathcal{O}(\Xi^{2}(x,p)),
\label{eq:consistency_general_final}
\end{equation}
demonstrating an $\mathcal{O}(\Xi(x,p))$ violation of $r = - 8 n_{t}$ even for a single canonical scalar when $E(N)$ evolves.

In the remainder, we apply these general results to four representative potentials. Throughout the model-by-model computations below, the fixed-$N$ mapping is obtained from Eq.  \eqref{eq:N_master_final}. Unless explicitly stated otherwise, we also adopt the constant-at-pivot approximation, at $N_{\ast}$,
\begin{equation}
E'(N_{\ast}) = U'(N_{\ast}) = 0,
\label{eq:const_pivot_final}
\end{equation}
so that the derivative-driven quantum contributions to $n_{s}$, $\alpha_{s}$, and $\beta_{s}$ at the pivot vanish and only the algebraic $(E_{\ast}, U_{\ast})$ corrections remain.

Let us now start with the analytic study  devoted to the quadratic power-law inflation in Section \ref{sec:QPL-infl}.

\subsection{Quadratic Power-Law Inflation}
\label{sec:QPL-infl}

Let us introduce the following canonical quadratic inflationary potential:
\begin{equation}
V(\varphi) = \frac{1}{2} m^{2} \varphi^{2}.
\label{eq:V_quad_final}
\end{equation}
For this model one immediately finds that $
V'/V = 2/\varphi$. The potential slow-roll parameters are given as
\bea
\begin{array}{ccl}
\epsilon_{V} &=& \frac{M_{\rm Pl}^{2}}{2}\left(\frac{2}{\varphi}\right)^{2}
 = \frac{2 M_{\rm Pl}^{2}}{\varphi^{2}}, \\
\eta_{V} &=& M_{\rm Pl}^{2} \frac{V''}{V}
 = \frac{2 M_{\rm Pl}^{2}}{\varphi^{2}}.
\end{array}
\label{eq:eps_eta_quad_final}
\eea
It is obvious that the inflation presumably ends when $\epsilon_{V}(\varphi_{\rm end}) = 1$, at which $\varphi^{2}$ is fixed to 
\begin{equation}
\varphi_{\rm end}^{2} = 2 M_{\rm Pl}^{2}.
\label{eq:phi_end_quad_final}
\end{equation}
Thus, the quantum corrected e-fold integral, Eq. \eqref{eq:N_master_final}, leads to
\bea
N & \simeq & \frac{1 + U_{\ast}}{M_{\rm Pl}^{2}} \int_{\varphi_{\rm end}}^{\varphi_{\ast}} \frac{V}{V'}\, d\varphi 
=
\frac{1 + U_{\ast}}{4 M_{\rm Pl}^{2}}
\left(\varphi_{\ast}^{2} - \varphi_{\rm end}^{2}\right).
\label{eq:N_quad_integral_final}
\eea
By solving this expression for $\varphi_{\ast}$ and using Eq. \eqref{eq:phi_end_quad_final}, one finds that 
\begin{equation}
\varphi_{\ast}^{2} = 4 M_{\rm Pl}^{2}\left(N + \frac{1}{2}\right) \frac{1}{1 + U_{\ast}} \simeq 4 M_{\rm Pl}^{2} \left(N + \frac{1}{2}\right) \left(1 - U_{\ast}\right)
+ \mathcal{O}(\Xi^{2}(x,p)).
\label{eq:phi_star_quad_final}
\end{equation}
The substitution of the preceding expression into Eq. \eqref{eq:eps_eta_quad_final} yields
\begin{equation}
\epsilon_{V\ast} = \eta_{V\ast}
 \simeq \frac{1 + U_{\ast}}{2 N + 1}.
\label{eq:eps_eta_star_quad_final}
\end{equation}

At fixed $N_{\ast}$, the standard slow-roll expressions are given as
\bea
\begin{array}{ccl}
n_{s} &\simeq & 1 - 6 \epsilon_{V\ast} + 2 \eta_{V\ast} = 1 - 4 \epsilon_{V\ast}, \\
r &\simeq & 16 \epsilon_{V\ast} \left(1 + \frac{3}{2} E_{\ast} + U_{\ast}\right),
\end{array}
\label{eq:ns_r_quad_start_final}
\eea
so that, the generic expressions read
\bea
\begin{array}{ccl}
n_{s}(N_{\ast}) & \simeq & 1 - \frac{4(1 + U_{\ast})}{2 N_{\ast} + 1}, \\
r(N_{\ast}) & \simeq & \frac{16}{2 N_{\ast} + 1}
\left(1 + \frac{3}{2} E_{\ast} + 2 U_{\ast}\right).
\end{array}
\label{eq:nsr_quad_final}
\eea
Assuming the constant-at-pivot condition, Eq. \eqref{eq:const_pivot_final}, the scalar spectral indexes, the running and the running-of-running are modified only through the background mapping so that
\bea
\begin{array}{ccl}
\alpha_{s}(N_{\ast}) &\simeq & - \frac{8(1 + U_{\ast})}{(2 N_{\ast} + 1)^{2}}, \\
\beta_{s}(N_{\ast}) &\simeq &
-\frac{32 (1 + U_{\ast})}{(2 N_{\ast} + 1)^{3}}.
\end{array}
\label{eq:alpha_beta_quad_final}
\eea

Starobinsky inflation is the second analytic study which is in outlined in Section \ref{sec:StarobinskyInfl}.

\subsection{Starobinsky Inflation}
\label{sec:StarobinskyInfl}

In scalar-tensor theories of gravity, when the gravitational action is expressed solely by the conventional Einstein--Hilbert term without any direct multiplication by scalar fields, the Einstein frame enables simplified field equations and allows for non-minimal coupling of matter to the metric \cite{EinsteinFrame}. The Starobinsky inflation is then described by \cite{Starobinsky:1979ty,doi:10.1142/9789814317344_0078}
\begin{equation}
V(\varphi) = \frac{3}{4} m^{2} M_{\rm Pl}^{2} \left(1 - e^{- a \varphi}\right)^{2}, 
\qquad a \equiv \sqrt{\frac{2}{3}} \frac{1}{M_{\rm Pl}}.
\label{eq:V_starob_final}
\end{equation}
Introducing $y \equiv e^{a \varphi}$, one derives the Starobinsky potential as well as first and second slow-roll quantities, respectively
\begin{equation}
\frac{V'}{V} = \frac{2 a}{y - 1},
\qquad
\epsilon_{V} = \frac{4/3}{(y - 1)^{2}},
\qquad
\eta_{V} = \frac{4}{3} \frac{2 - y}{(y - 1)^{2}}.
\end{equation}
The quantum corrected $N - y_{\ast}$ relationship follows from
\begin{equation}
N \simeq \frac{3}{4}(1 + U_{\ast})
\left[y_{\ast} - y_{\rm end} - \ln\!\left(\frac{y_{\ast}}{y_{\rm end}}\right)
\right],
\label{eq:N_starob_relation_final}
\end{equation}
where $\epsilon_{V}(y_{\rm end}) = 1$ presumably fixes $y_{\rm end} = 1 + \frac{2}{\sqrt{3}}$. For $N \gg 1$, this reduces to $y_{\ast} \simeq \frac{4 N}{3} \left(1 - U_{\ast}\right)$. Consequently, one derives the first and second slow-roll parameters, 
\begin{equation}
\epsilon_{V\ast} \simeq \frac{3}{4 N^{2}}(1 + 2 U_{\ast}),  \qquad \eta_{V\ast} \simeq - \frac{1 + U_{\ast}}{N}.
\end{equation}
Then, the sets of observable $n_s$ and $r$ read
\bea
\begin{array}{ccl}
n_{s} & \simeq & 1 - \frac{2}{N}(1 + U_{\ast}), \\
r & \simeq & \frac{12}{N^{2}} \left(1 + \frac{3}{2} E_{\ast} + 2 U_{\ast}\right),
\end{array}
\eea
with the scalar spectral indexes
\bea
\begin{array}{ccl}
\alpha_{s} &\simeq & - \frac{2(1 + U_{\ast})}{N^{2}}, \\
\beta_{s} &\simeq & \frac{4(1 + U_{\ast})}{N^{3}}.
\end{array}
\eea

A third analytic study is devoted to d-brane (Coulomb) inflation in Section 
\ref{sec:dBrane}

\subsection{D-Brane (Coulomb) Inflation}
\label{sec:dBrane}

In order to establish an effective Coulomb potential, let us now analyze d-brane (Coulomb) inflation
\cite{PhysRevD.76.103512}, which is influenced by the potential energy arising from the interaction between a d3-brane and an anti-d3-brane, that based on cosmological string-theory models,
\begin{equation}
V(\varphi) = V_{0} \left(1 - \frac{\mu^{p}}{\varphi^{p}}\right),
\qquad
p>0,
\qquad
\varphi^{p} \gg \mu^{p},
\label{eq:V_dbrane_final}
\end{equation}
where $p$ is the power of the Coulomb term which represents the dimensionality of the branes and the geometry of the extra dimensions. Here, $\mu$ is a constant with dimensions of mass. It encodes the strength of the brane-antibrane Coulomb attraction. To leading order, one finds that 
\begin{equation}
\varphi_{\ast}^{p + 2} \simeq
p(p + 2) \mu^{p} M_{\rm Pl}^{2} N \left(1 - U_{\ast}\right),
\label{eq:phi_star_dbrane_final}
\end{equation}
Then, the slow-roll parameters are given as
\bea
\begin{array}{ccl}
\eta_{V\ast} &\simeq & - \frac{p + 1}{p + 2}\frac{1 + U_{\ast}}{N}, \\
\epsilon_{V\ast} &\simeq & \frac{p^{2}}{2(p + 2)^{2}} \frac{1 + U_{\ast}}{N^{2}}.
\end{array}
\eea
Accordingly, one obtains
\bea
\begin{array}{ccl}
n_{s}(N_{\ast}) & \simeq &
1 - \frac{2(p + 1)}{p + 2} \frac{1 + U_{\ast}}{N_{\ast}}, \\
r(N_{\ast}) & \simeq &
\frac{8 p^{2}}{(p + 2)^{2}} \frac{1}{N_{\ast}^{2}} \left(1 + \frac{3}{2} E_{\ast} + 2 U_{\ast}\right),
\end{array}
\eea
with the scalar spectral indexes
\bea
\begin{array}{ccl}
\alpha_{s} &\simeq & - \frac{2(p + 1)}{p + 2}\frac{1 + U_{\ast}}{N^{2}}, \\
\beta_{s} &\simeq & \frac{4(p + 1)}{p + 2}\frac{1 + U_{\ast}}{N^{3}}.
\end{array}
\eea

Now, we arrive at the fourth analytic study based on natural inflation, Section \ref{sec:NatInfl}.

\subsection{Natural Inflation}
\label{sec:NatInfl}

In natural inflation, the inflaton is identical with a pseudo-Nambu--Goldstone boson. An approximate shift symmetry is introduced  preserve the inflation potential from steepening, allowing slow‑roll inflation to occur naturally without fine‑tuning \cite{Freese:2004un}
\begin{equation}
V(\varphi) = \Lambda^{4} \left[1 + \cos\!\left(\frac{\varphi}{f}\right)\right].
\label{eq:V_natural_final}
\end{equation}
While $\Lambda$ sets the energy scale, $f$ is  an axion decay constant which sets the width of the cosine potential so that larger $f$ makes the potential flatter (successful inflation) while smaller $f$ defines a steeper potential (no slow-roll). Defining $\theta = \varphi/f$, one finds that
\bea
\begin{array}{ccl}
\epsilon_{V} &=& \frac{M_{\rm Pl}^{2}}{2 f^{2}} \tan^{2}\!\left(\frac{\theta}{2}\right), \\
\eta_{V} &=& - \frac{M_{\rm Pl}^{2}}{f^{2}}
 \frac{\cos(\theta)}{1 + \cos(\theta)}.
 \end{array}
\eea
The exact quantum corrected $N$-$\theta_{\ast}$ relationship is given as
\begin{equation}
\sin\!\left(\frac{\theta_{\ast}}{2}\right)
= \sin\!\left(\frac{\theta_{\rm end}}{2} \right) \exp\! \left[ - \frac{N M_{\rm Pl}^{2}}{2 f^{2}} \left(1 - U_{\ast}\right) \right],
\label{eq:theta_star_natural_final}
\end{equation}
where $\tan (\theta_{\rm end}/2) = \sqrt{2} f/M_{\rm Pl}$.
The sets of observable read
\bea
\begin{array}{ccl}
n_{s} &=& 1 - \frac{3 M_{\rm Pl}^{2}}{f^{2}} \tan^{2}\!\left(\frac{\theta_{\ast}}{2}\right) - \frac{2 M_{\rm Pl}^{2}}{f^{2}}
 \frac{\cos(\theta_{\ast})}{1 + \cos(\theta_{\ast})}, \\
r &=& 8 \frac{M_{\rm Pl}^{2}}{f^{2}}
\tan^{2}\! \left(\frac{\theta_{\ast}}{2}\right)
\left(1 + \frac{3}{2} E_{\ast} + 2 U_{\ast}\right).
\end{array}
\eea

Having derived the analytical expressions from different inflation models, Section \ref{sec:AnltcReslt}, we can now proceed with our numerical analysis, Section \ref{sec:NmrcReslt}.

\section{Numerical Results} 
\label{sec:NmrcReslt}

\begin{figure}
\centering
\includegraphics[width=\textwidth]{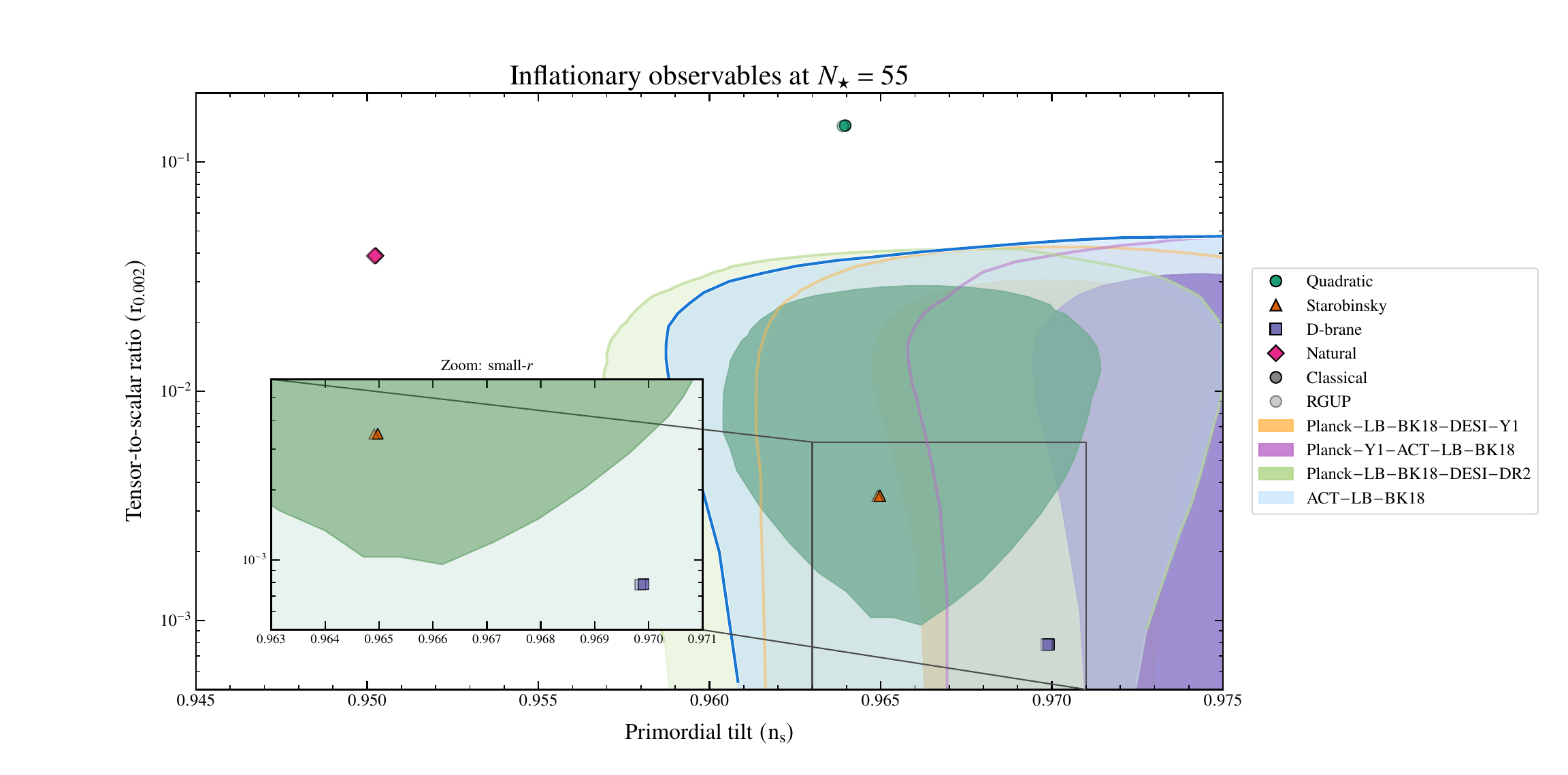}
\caption{For the various inflation models, Section \ref{sec:AnltcReslt}, the tensor-to-scalar ratio $r$ is depicted as a function of the primordial tilt $n_s$, at fixed $N_{\ast}=55$. Opaque symbols denote classical results, while translucent ones marker quantum corrected predictions for $(E_{\ast}, U_{\ast})=(- 6  \times 10^{-3},\,2 \times 10^{-3})$.
}
\label{fig:nsr_fixedN}
\end{figure}

\begin{figure}
\centering
\includegraphics[width=\textwidth]{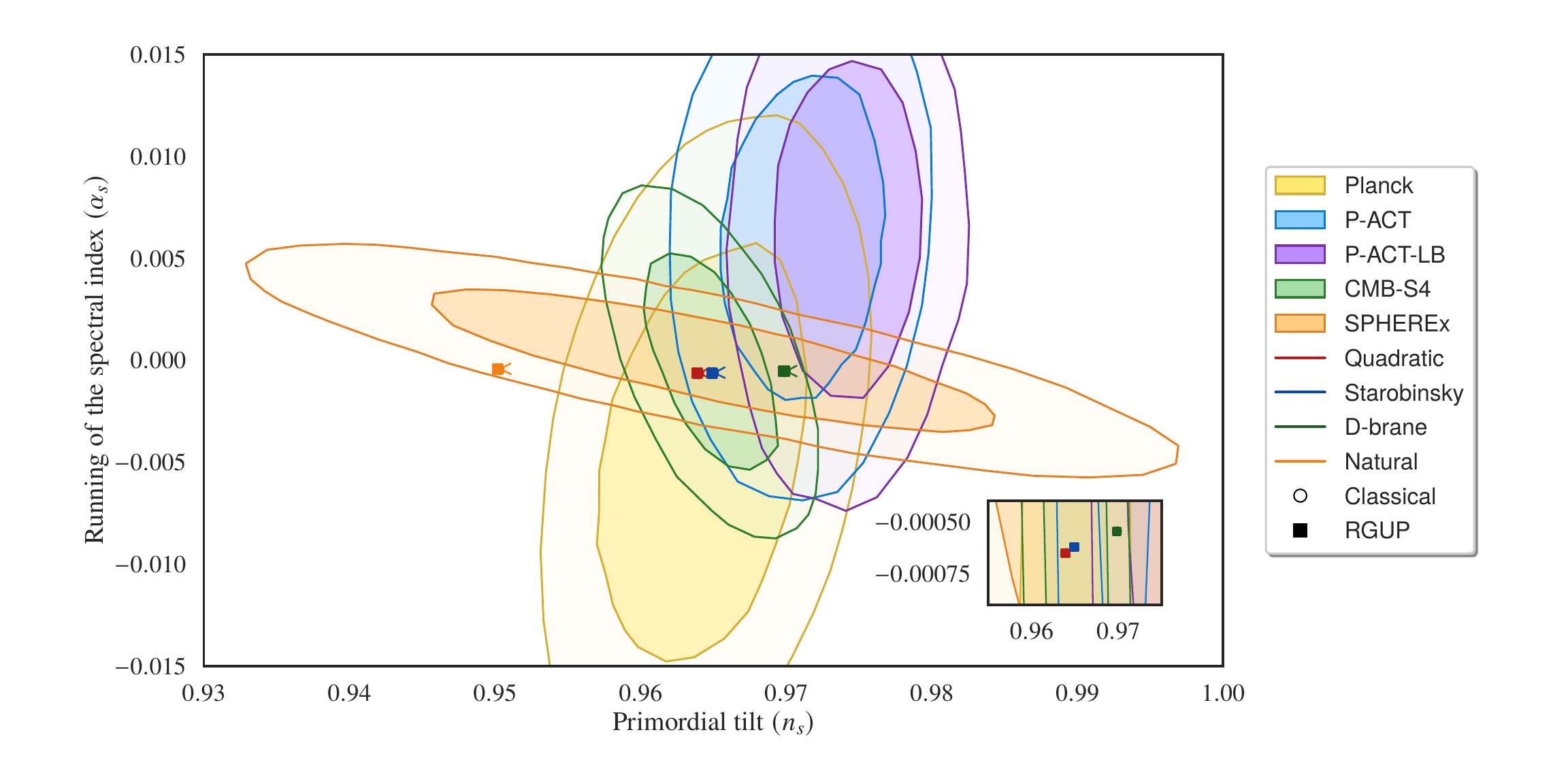}
\caption{The same as in Fig. \ref{fig:nsr_fixedN}, but here the running of the special index $\alpha_s$ is presented in dependence on the primordial tilt $n_s$.
Colored contours show current and forecast CMB constraints \cite{Bianchini:2025dhf,PhysRevD.94.023523} such as Planck \cite{Patel:2025nxh,Jense:2025wyg}, LiteBIRD \cite{LiteBIRD:2025nom,Petretti:2024mjy}, and CMB-S4 \cite{Bianchini:2025dhf,CMB-S4:2024zgz,Mehta:2024sye}.  Open circular markers denote classical slow-roll predictions, while filled square markers correspond to the results obtained from the proposed geometric quantization for the benchmark values $(E_{\ast},U_{\ast})=(-6 \times 10^{-3},\,2 \times 10^{-3})$ and  $E'(N_{\ast})=U'(N_{\ast})=0$.
Thin arrow heads indicate classical and quantum points, explicitly referring to the direction and magnitude of the quantum-geometric shift at fixed $N_\star$. 
}
\label{fig:ns_alpha_fixedN}
\end{figure}

At e-fold number $N_{\ast}=55$, the results obtained from the various types of the inflationary observables are presented in Fig.~\ref{fig:nsr_fixedN} and \ref{fig:ns_alpha_fixedN}. In Fig.~\ref{fig:nsr_fixedN}, we compare the four representative inflationary potentials discussed in Section \ref{sec:AnltcReslt}. Additionally, we present numerical results derived from classical models. Specifically, for a fixed value of $N_{\ast}=55$, the tensor-to-scalar ratio $r$ is illustrated as a function of the primordial tilt $n_s$. The translucent markers represent quantum corrected predictions for $(E_{\ast}, U_{\ast})=(- 6 \times 10^{-3},\,2 \times 10^{-3})$, while the opaque symbols indicate classical results. In inflation models, quantum deformation seems to cause a small but consistent change in the $(n_s, r)$ plane, moving the points toward a lower tensor-to-scalar ratio $(r)$ and a slightly smaller scalar tilt $(n_s)$. The main effect is the universal decrease in $r$, which matches the first-order correction formula $r \simeq 16 \epsilon_V* (1 + 1.5 E* + U*)$. This correction comes from quantum effects and how they change the gravity part of the model. We also find that the change in $n_s$ is small and depends on the specific model. It happens indirectly through the way $U_{\ast}$ relates the number the Universe has expanded, the so-called e-folds, to when certain fluctuations leave the horizon, the so-called horizon-exit. When the derivatives $E'(N_{\ast})$ and $U'(N_{\ast})$ are zero, the corrections that depend on scale disappear. The small inset shows the region with low $r$, which is important for Starobinsky and d-brane inflation models. In this region, the corrections from the proposed geometric quantization approach are very small and stay within what we can observe. Overall, the numerical results in Fig. \ref{fig:nsr_fixedN} show that the geometric quantization approach keeps the order of different inflation models the same in the $(n_s, r)$ graph. It also slightly lowers both the tensor-to-scalar ratio $r$ and the distance the inflaton travels during inflation, matching the numbers in Tab. \ref{tab:QC_big_table_clean}.

The results summarized in Fig.~\ref{fig:ns_alpha_fixedN} quantify the leading-order impact of geometric quantum corrections on inflationary observable evaluated at a fixed e-fold number, $N_\star=55$. For all four inflationary potentials considered, Section~\ref{sec:AnltcReslt}, the geometric quantum deformation induces systematic but parametrically small shifts in the scalar spectral index $n_s$ and its running $\alpha_s$, controlled by the background scalars $(E_{\ast},U_{\ast})$. Imposing the constant-at-pivot condition, $E'(N_{\ast})=U'(N_{\ast})=0$, eliminates all explicit derivative contributions to the spectral indexes, so that the remaining corrections arise solely through the $U_{\ast}$-dependent modification of the horizon-exit field value entering the fixed-$N_{\ast}$ mapping. Consequently, the geometric quantum corrections to $n_s$ and $\alpha_s$ are subleading and model dependent, typically of order $\mathcal{O}(10^{-5}$--$10^{-4})$, in precise agreement with the first-order analytic expressions in Eqs.~\eqref{eq:ns_general_final} and \eqref{eq:alpha_beta_general_final} and with the numerical results reported in Tab.~\ref{tab:QC_big_table_clean}. 

A salient feature of these results is the universality of the geometric quantum correction structure across different inflationary potentials. At fixed $N_{\ast}$, the ordering of inflation models in the $(n_s,\alpha_s)$ plane seems to be preserved, indicating that geometric quantum effects act primarily as a controlled deformation of classical slow-roll predictions rather than as a mechanism for qualitative model discrimination. This behavior follows directly from the structure of the first-order RGUP expansion, in which the scalar amplitude depends on the combination $-\tfrac12 E_{\ast}-U_{\ast}$, while the spectral tilt and its running receive corrections only through background reparametrization effects. As a result, plateau-type models such as Starobinsky and d-brane inflation remain consistent with current observational constraints, whereas quadratic and natural inflation models continue to lie in comparatively less constrained regions. Overall, the numerical analyses demonstrate that quantum induced corrections yield theoretically well-defined and observationally mild deviations from classical single-field slow-roll inflation, providing a consistent and predictive framework for incorporating quantum phase-space effects into inflationary cosmology. 

\begin{figure}[t]
\centering
\includegraphics[width=\textwidth]{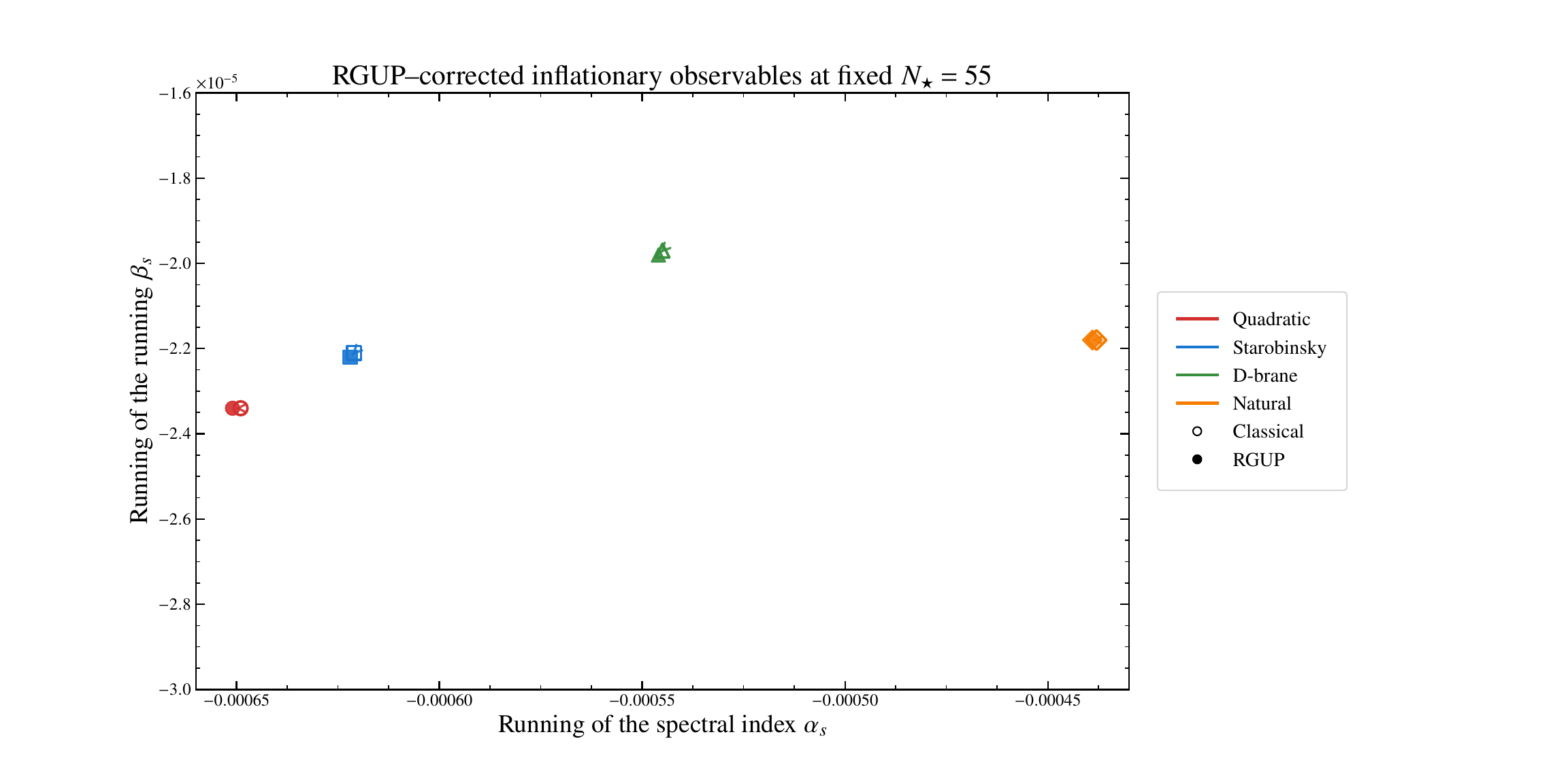}
\caption{The same as in Fig. \ref{fig:ns_alpha_fixedN} but here for running-of-running $\beta_s$ against running of the spectral index $\alpha_s$. Open markers denote classical results, while filled markers correspond to quantum-corrected values for the benchmark $(E_{\ast}, U_{\ast}) = (- 6 \times 10^{-3},\, 2 \times 10^{-3})$ with $E'(N_{\ast})=U'(N_{\ast})=0$. 
\label{fig:alpha_beta_fixedN}
}
\end{figure}

Figure~\ref{fig:alpha_beta_fixedN} subtracts the higher-derivative structure of the scalar spectrum by displaying the running of the spectral index $\alpha_s$ and the running-of-running $\beta_s$ for the four representative potentials, at fixed $N_{\ast}=55$. A first key observation is that all models appear to cluster in the theoretically expected slow-roll regime, with $\alpha_s < 0$ and $\beta_s < 0$ at magnitudes $\sim 10^{-4}$ and $\sim 10^{-5}$, respectively, confirming that the spectra remain perturbatively close to scale invariance even after the quantum deformation. Second, the geometric quantum correction acts as a controlled, model-dependent perturbation rather than a significant modification of the predictions: the quantum-deformed remain very close to their classical counterparts, indicating that at the chosen benchmark pivot, the quantum geometry primarily renormalizes the background mapping at fixed $N_{ ext{ast}}$ rather than creating new scale dependence. This behavior is consistent with the constant-at-pivot assumption $E'(N_{\ast})=U'(N_{\ast})=0$, under which the explicit derivative-driven contributions to $n_s$, $\alpha_s$, and $\beta_s$ vanish in the general expressions, and any residual shift is inherited indirectly through the $U_{\ast}$-modified horizon-exit solution for $\varphi_{\ast}(N_{\ast})$. Accordingly, the displacements are of order $\mathcal{O}(10^{-6}$ - $10^{-5})$ in $(\alpha_s, \beta_s)$ plane, matching the numerical differences reported in Tab.~\ref{tab:QC_big_table_clean}.   Finally, the relative separation among the various inflation models is dominated by the underlying potential shape, e.g., plateau-like versus large-field behavior, while the geometric quantum corrections preserve this ordering, indicating that the primary discriminant in the $(\alpha_s, \beta_s)$ plane remains the classical slow-roll dynamics and that the geometric quantum effects enter as a small, theoretically well-controlled deformation at fixed $N_{\ast}$. 

\begin{table}[t]
\centering
\resizebox{\textwidth}{!}{%
\begin{tabular}{ | l | c | c c c c c | }
\hline
Model & Approach & $n_s$ & $r$ & $\alpha_s$ & $\beta_s$ & $\Delta\varphi/M_{\rm Pl}$ \\
\hline \hline
\multirow{2}{6em}{Quadratic} & Classical & 0.96396 & 0.1441 & $-6.49\times10^{-4}$ & $-2.34\times10^{-5}$ & 13.485 \\
& Quantum & 0.96389 & 0.1434 & $-6.51\times10^{-4}$ & $-2.34\times10^{-5}$ & 13.471 \\
\hline
\multirow{2}{6em}{Starobinsky} & Classical & 0.96498 & $3.50\times10^{-3}$ & $-6.21\times10^{-4}$ & $-2.21\times10^{-5}$ & 4.413 \\
& Quantum & 0.96491 & $3.49\times10^{-3}$ & $-6.22\times10^{-4}$ & $-2.22\times10^{-5}$ & 4.410 \\
\hline
\multirow{2}{6em}{D-Brane} & Classical & 0.96991 & $7.86\times10^{-4}$ & $-5.45\times10^{-4}$ & $-1.97\times10^{-5}$ & 1.996 \\
& Quantum & 0.9698 & $7.83\times10^{-4}$ & $-5.46\times10^{-4}$ & $-1.98\times10^{-5}$ & 1.995 \\
\hline
\multirow{2}{6em}{Natural} & Classical & 0.95025 & $3.90\times10^{-2}$ & $-4.38\times10^{-4}$ & $-2.18\times10^{-5}$ & 10.944 \\
& Quantum & 0.9502 & $3.89\times10^{-2}$ & $-4.39\times10^{-4}$ & $-2.18\times10^{-5}$ & 10.937 \\
\hline
\end{tabular}
}
\caption{A comparison between various inflationary observable evaluated at fixed e-fold number $N_\star=55$. For each of the inflation models considered, the classical observation is compared with the quantum-deformed one. The quantum benchmark corresponds to $(E_{\ast},U_{\ast})=(- 6 \times 10^{-3},\, 2 \times 10^{-3})$ with $E'(N_{\ast})=U'(N_{\ast})=0$.
For the d-brane model $p=4$ and $\mu=M_{\rm Pl}$, while for natural inflation $f=5 M_{\rm Pl}$.
}
\label{tab:QC_big_table_clean}
\end{table}

\section{Conclusions and outlook}
\label{sec:conclusions}

The geometric quantization framework developed in this work provides a covariant and dimensionally consistent mechanism for incorporating quantum-geometric corrections into inflationary cosmology. Quantum effects are introduced through a controlled rank-$2$ deformation of the background metric sourced by momentum-space structure, without violating general covariance or introducing additional propagating degrees of freedom. All results discussed follow directly from the quantum-deformed geometric construction of the fundamental metric, which can be reduced straightforwardly to their classical counterparts in the limits $\Xi(p_0)\to0$ and $\epsilon\to0$.

In this work, we have developed a covariant and systematically controlled framework for computing a set of inflationary observable in the presence of quantum-induced corrections. Starting from a phase-space formulation of gravity with a momentum-dependent deformation of the spacetime metric, we have shown how quantum-gravitational effects can be consistently propagated from the underlying geometric structure to measurable cosmological quantities, without introducing additional fields, explicit Lorentz violation, or {\it ad hoc} modifications of the inflaton sector.

A central methodological choice of this analysis is the evaluation of observable at fixed e-fold number and the restriction to a strict first-order expansion in the geometric quantum deformation parameter $\Xi$. Within this regime, we derived a closed and internally consistent set of expressions for the scalar and tensor power spectra, their spectral tilts, running, and the tensor-to-scalar ratio. All quantum corrections are organized in terms of two dimensionless background functions, $E(N)$ and $U(N)$, arising from distinct covariant contractions of the momentum-frame field. The function $U(N)$ governs the reparametrization of the background inflaton dynamics and the mapping between field space and e-folds, while $E(N)$ controls the effective geometric measure relevant for tensor perturbations and gravitational-wave amplitudes. We conclude that at fixed $N_{\ast}$ and under the condition $E'(N_{\ast})=U'(N_{\ast})=0$, the dominant geometric quantization impact is a universal attenuation of the tensor amplitude through $r \simeq 16 \epsilon_{V\ast}(1+\tfrac32 E_\ast+U_\ast)$, together with a smaller, model-dependent shift of $(n_s,\alpha_s,\beta_s)$ induced indirectly by the $U_{\ast}$-modified horizon-exit mapping. For the benchmark $(E_{\ast},U_{\ast})=(- 6 \times 10^{-3}, 2 \times 10^{-3})$, all four inflationary potentials exhibit a small decrease in $r$, at the $\sim 10^{-3}$-$10^{-2}$ relative level, while $n_s$ shifts downward by $\mathcal{O}(10^{-4})$-$\mathcal{O}(10^{-5})$ depending on the model. The field excursion $|\Delta \varphi|/M_{\rm Pl}$ is also reduced for $U_{\ast}>0$ in each case, consistent with the fixed-$N$ background reparameterization. 

The numerical analyses reveals that all four models share the same geometric quantization features, at fixed $N_{\ast}$. The background mapping is modified exclusively through $U_{\ast}$, the scalar amplitude depends on $-\tfrac12 E_{\ast}-U_{\ast}$, and the tensor amplitude depends only on $E_{\ast}$. Consequently, one estimates that
\begin{equation}
r(N_{\ast}) = 16 \epsilon_{V\ast}(N_{\ast}) \left(1 + \frac{3}{2} E_{\ast} + U_{\ast}\right) + \mathcal{O}(\Xi^{2}(x,p)).
\end{equation}
Thus, one concludes that any geometric quantum induced scale dependence arises from derivatives $E'(N)$ and $U'(N)$.

In summary, the quantum-geometric corrections introduced in this work seem to admit a clear physical interpretation in terms of measure rescaling and momentum-frame-induced kinetic deformations. These effects modify inflationary observables in a controlled and predictive manner, preserve the classical limit, and provide a well-defined phenomenological window onto possible quantum-gravitational structure in the early Universe.

\section*{Conflicts of Interest}

The authors declare that there are no conflicts of interest regarding the publication of this published article!

\section*{Dataset Availability}

All data generated or analyzed during this study are included in this published article. All of the material is owned by the authors.

\section*{Author contributions}
The responsibility for proposing the conception of the present study lies with AT, who also undertook the tasks of designing and managing the research, interpreting the results, and preparing the manuscript. SKS was responsible for deriving the expressions and proposing physical interpretations. SOA, AAA and MN contributed to the writing and proofreading of the manuscript. The final version of the manuscript was unanimously approved by all authors.

\section*{Funding}

The authors declare that this research received no specific grants from any funding agency in the public, commercial, or not-for-profit sectors.

\section*{Competing interests}

The authors confirm that there are no relevant financial or non-financial competing interests to report.

\bibliographystyle{unsrtnat}
\bibliography{2025-03-05-RevisitedPowerLaw1}

\newpage

\appendix

\renewcommand{\thefigure}{A.\arabic{figure}} 
\setcounter{figure}{0} 
\setcounter{table}{0} 
\renewcommand{\thetable}{A\arabic{table}} 

\section{Physical Interpretation of  Quantum-Deformed Inflationary Observables}
\label{app:physical_interpretation}

Substituting the quantum-corrected geometric quantities into the standard inflationary observables yields modified expressions for the primordial spectra and their derived parameters. To leading order, these expressions take the following schematic forms:
\begin{align}
P_{\mathcal R}(k) &\simeq P_{\mathcal R}^{(0)}(k) \left[1 - \frac{1}{2}\, \epsilon - \Xi(p_0) \frac{(\dot p_0^0)^2}{\mathcal F^2}\right], \\
P_h(k) &\simeq \frac{2}{\pi^2}\frac{H^2}{M_{\rm Pl}^2} \left(1 + \epsilon\right), \\
n_s-1 &\simeq \frac{d\ln P_{\mathcal R}^{(0)}(k)}{d\ln k} + \frac{d}{d\ln k}\left[- \frac{1}{2}\,\epsilon - \Xi(p_0)\frac{(\dot p_0^0)^2}{\mathcal F^2} \right], \\
r &\simeq r^{(0)} \left[1 + \frac{3}{2}\,\epsilon
+ \Xi(p_0)\frac{(\dot p_0^0)^2}{\mathcal F^2}
\right], \\
\alpha_s &\simeq \alpha_s^{(0)} + \frac{d^2}{d(\ln k)^2} \left[- \frac{1}{2}\,\epsilon - \Xi(p_0)\frac{(\dot p_0^0)^2}{\mathcal F^2}\right].
\end{align}
These expressions illustrate how quantum-geometric effects modify inflationary observables through a small number of dimensionless, geometrically motivated quantities. Let us now summarize the physical interpretation of the key ingredients entering the previous corrections.
\begin{itemize}
\item \textbf{Metric and measure corrections.}  
The quantum deformation of the spacetime metric induces a modification of the invariant volume element,
\bea
\sqrt{-\widetilde g} &=& \sqrt{-g} \left(1+\tfrac12\epsilon\right), \nn
\eea
where $\epsilon$ is a dimensionless measure correction arising from the quantum geometry of phase space. This term can be interpreted as an effective rescaling of the gravitational sector, analogous to a background expectation value of metric fluctuations. It affects both the background dynamics and the normalization of perturbations, while remaining perturbatively small and vanishing in the classical limit.
\item \textbf{Effective kinetic normalization.}  
In the presence of a momentum-dependent inverse-metric deformation, the effective kinetic term of the homogeneous inflaton receives additional corrections beyond the measure rescaling. In particular, the time-like contraction
\bea
-\Xi(p_0)\frac{(\dot p_0^0)^2}{\mathcal F^2}, \nn
\eea
modifies the normalization of the kinetic energy density. The combined effect of measure and inverse-metric corrections can be summarized by an effective kinetic deformation $\delta_K$, which rescales the Mukhanov--Sasaki variable \cite{Gomar:2014faa,CastelloGomar:2016rjj} as $z\to z(1+\delta_z)$. Since the scalar power spectrum scales as $P_{\mathcal R}\propto z^{-2}$, even small values of $\delta_z$ can lead to observable deviations from the predictions of classical single-field inflation.

\item \textbf{Scalar versus tensor perturbations.}  
Tensor modes are primarily sensitive to the measure deformation $\epsilon$, which rescales the normalization of gravitational-wave fluctuations, leading to $\Delta_t\simeq\epsilon$. Scalar modes, by contrast, are affected by both the measure correction and the modified kinetic normalization, resulting in
\bea
\Delta_s &\simeq &  - \frac12\,\epsilon -  \Xi(p_0)\frac{(\dot p_0^0)^2}{\mathcal F^2}. \nn
\eea
This asymmetry between scalar and tensor sectors is a characteristic feature of the quantum-geometric framework and directly impacts the tensor-to-scalar ratio. In particular, it provides a geometric mechanism for deviations from the classical single-field consistency relation $r=-8n_t$.

\item \textbf{Spectral observables and consistency relations.}  
The scalar spectral index acquires a correction $\delta n_s=d\Delta_s/d\ln k$, governed by the scale dependence of $\epsilon$ and the momentum-frame contraction. The tensor-to-scalar ratio is modified by
\bea
\Delta_t - \Delta_s &=& \frac{3}{2}\, \epsilon + \Xi(p_0)\frac{(\dot p_0^0)^2}{\mathcal F^2}, \nn
\eea
which may be constrained by current and forthcoming CMB constraints \cite{Bianchini:2025dhf,PhysRevD.94.023523} such as Planck \cite{Patel:2025nxh,Jense:2025wyg}, LiteBIRD \cite{LiteBIRD:2025nom,Petretti:2024mjy}, and CMB-S4 \cite{Bianchini:2025dhf,CMB-S4:2024zgz,Mehta:2024sye}. A detection of departures from the classical consistency relation would therefore provide indirect evidence for quantum-geometric effects encoded in the phase-space structure of spacetime.

\item \textbf{Scale dependence and phenomenology.}  
While $\epsilon$ and $\Xi(p_0)(\dot p_0^0)^2/\mathcal F^2$ are treated as approximately constant at leading order, mild evolution with time or scale can generate additional contributions to the running and running-of-running of the scalar spectral index. Such effects offer a potential observational discriminant between quantum-corrected inflationary scenarios and their classical counterparts, and may become accessible through high-precision measurements of the CMB power spectrum, higher-point correlations, or primordial gravitational waves.

\end{itemize}

\end{document}